\documentclass{pnastwo}

\usepackage[dvips]{graphicx}
\usepackage{pnastwof}
\usepackage{amssymb,amsfonts,amsmath}
\usepackage[normalem]{ulem}
\usepackage{color}

\url{www.pnas.org/cgi/doi/10.1073/pnas.0709640104}
\copyrightyear{2008} \issuedate{17-07-2012} \volume{109}
\issuenumber{Issue Number}

\begin{document}

\title{Imaging electronic quantum motion with light}

\author{Gopal Dixit\affil{1}{Center for Free-Electron Laser Science, DESY,
            Notkestrasse 85, D-22607 Hamburg, Germany},
Oriol Vendrell\affil{1}{}, \and Robin
Santra\affil{1},\affil{2}{Department of Physics, University of
Hamburg, D-20355 Hamburg, Germany }}

\contributor{Submitted to Proceedings of the National Academy of Sciences
of the United States of America}

\maketitle

\begin{article}
\begin{abstract}
Imaging the quantum motion of electrons not only in real-time, but
also in real-space is essential to understand for example bond
breaking and formation in molecules, and charge migration in
peptides and biological systems. Time-resolved imaging
interrogates the unfolding electronic motion in such systems. We
find that scattering patterns, obtained by X-ray time-resolved
imaging from an electronic wavepacket, encode spatial and temporal
correlations that deviate substantially from the common notion of
the instantaneous electronic density as the key quantity being
probed. Surprisingly, the patterns provide an unusually visual
manifestation of the quantum nature of light. This quantum nature
becomes central only for non-stationary electronic states and has
profound consequences for time-resolved imaging.
\end{abstract}

\keywords{X-ray imaging | quantum electrodynamics | electronic
wavepacket}

\abbreviations{TRI, time-resolved imaging; QED, quantum
electrodynamics; DSP, differential scattering probability}

The scattering of light from matter is a fundamental phenomenon
that is widely applied to gain insight about the structure of
materials, biomolecules and nanostructures.
The wavelength of X-rays is of the order of atomic distances in
liquids and solids, which makes X-rays a very convenient probe for
obtaining real-space, atomic-scale structural information of
complex materials, ranging from molecules~\cite{Ihee} to
proteins~\cite{Chapman} and viruses~\cite{Seibert}. The power of
X-ray scattering relies as well on the fact that the X-rays
interact very weakly with the electrons in matter. In a given
macroscopic sample, generally no more than one scattering event
per X-ray photon takes place, the probability for multiple
scattering being extremely small. The key quantity in \mbox{X-ray}
scattering is the differential scattering probability (DSP), which
is related to the Fourier transform of the electronic density
$\rho(\mathbf{x})$ as follows~\cite{ashcroft, james1982}

\begin{equation}\label{eq1}
\frac{dP}{d\Omega} = \frac{dP_{e}}{d\Omega} \left|\int d^{3}x\;
\rho(\mathbf{x}) ~ e^{i {\mathbf{Q} \cdot \mathbf{x}}}
\right|^{2},
\end{equation}
where $\frac{dP_{e}}{d\Omega}$ is the differential scattering
probability  from a free electron and $\mathbf{Q}$ is proportional
to the momentum transfer of the scattered light. Procedures exist,
for both crystalline~\cite{Hauptman} and non-crystalline
samples~\cite{Miao}, to reconstruct $\rho(\mathbf{x})$ from the
scattering pattern.
Equation~(\ref{eq1}) can be obtained from a purely classical
description of electromagnetic radiation scattered by a stationary
electron density~\cite{james1982}, yielding a result identical to
that obtained from a quantum electrodynamics (QED) description of
light.

Equation~(\ref{eq1}) gives us access to a static view of the
electronic density. On the other hand, much progress has been made
in recent years towards understanding electronic motion with
time-domain table-top experiments~\cite{Haessler, Tzallas,
Hockett, Niikura, Goulielmakis, Sansone}, owing to the
availability of laser pulses on the sub-fs time
scale~\cite{Goulielmakis1, Hentschel}.
An ultimate goal of imaging applications encompasses unraveling
the motion of electrons and atoms with spatial and temporal
resolutions of order 1~{\AA} and 1~fs, respectively~\cite{Krausz}.
Recent breakthroughs make it possible to generate hard
\mbox{X-ray} pulses of a few fs~\cite{Emma, Pile}, and pulses of
length 100~as can in principle be realized~\cite{Emma1, Zholents}.
Hence, a fundamental question that needs to be addressed is: How
does an ultrashort light pulse interact and scatter from a
non-stationary quantum system?

Figure~1 shows schematically how the real-time, real-space
dynamics of an electronic wave-packet can be probed by scattering
a short X-ray pulse from it. In the example selected here, the
electronic wavepacket is a coherent superposition of eigenstates
of atomic hydrogen, with a wavepacket oscillation period of T =
$6.25$ fs. The electronic dynamics of the wavepacket is probed as
a function of pump-probe delay time $t$. Previously, a similar
example was used to illustrate time resolved imaging of an
electronic wavepacket using ultrafast electron
diffraction~\cite{Starace}. There, the instantaneous electron
density was assumed to be the main quantity being imaged.

In order to describe TRI, it is tempting to apply a similar
treatment as used to obtain Eq.~(\ref{eq1}), i.e., a classical
description of the \mbox{X-ray} pulse being scattered from a
non-stationary electronic system  described by the wavepacket.
Therefore, under the assumption that the \mbox{X-ray} pulse
duration is shorter than the time scale on which the motion of the
electronic wavepacket unfolds (see supplementary information), the
expression for the DSP becomes
\begin{equation}\label{eq11}
\frac{dP}{d\Omega} = \frac{dP_{e}}{d\Omega} \left|\int d^{3}x\;
\rho(\mathbf{x}, t) ~ e^{i\mathbf{Q \cdot x}} \right|^{2}.
\end{equation}
According to Eq.~(\ref{eq11}), TRI would be expected to provide
access to the instantaneous electron density, $\rho(\mathbf{x},
t)$, as a function of the delay-time $t$. The idea to assign an
additional degree of freedom to the electronic density in
Eq.~(\ref{eq1}), by replacing $\rho(\mathbf{x})$ with
$\rho(\mathbf{x}, t)$, is not new, and it has already been
proposed and used in the recent past~\cite{Krausz, Neutze, Quiney,
Jurek}.
However, a more careful consideration of the physics behind a
time-resolved scattering process, quickly points to pitfalls in
the simple logic underlying Eq.~(\ref{eq11}).
Equation~(\ref{eq11}) implies that the scattering of light does
not affect the properties of the wavepacket. But in order to image
quantum motion on an ultrafast time scale, one needs an ultrashort
pulse, which has an unavoidable spectral bandwidth. Due to the
finite bandwidth of an ultrashort pulse, it is energetically
impossible to detect whether or not in the scattering process a
transition between the eigenstates spanning the wavepacket, or
transitions to other states closer in energy than the pulse
bandwidth, has taken place. As a result, a physically correct
treatment of ultrafast imaging from the electronic wavepacket will
necessarily allow for transitions among electronic states caused
by the scattering process (Compton type process). In other words,
it must be expected that light scattering changes the wavepacket
being imaged, and that this effect will be reflected in the image.
In light of this, it is questionable whether Eq.~(\ref{eq11}) is
at all justified.

In the following, we apply a consistent QED description of light
and a quantum mechanical description of matter. Only in this way
can electronic transitions during the scattering event be properly
accounted for~\cite{Henriksen, Tanaka}. We find that the photon
concept, i.e. a quantum of light scattering from the electronic
system, is crucial in the description of TRI. For a sufficiently
short, coherent pulse, the resulting expression for the DSP is
(see supplementary information)
\begin{equation}\label{eq2}
\frac{dP}{d\Omega} = \frac{dP_{e}}{d\Omega} \int d^{3}x \int
d^{3}x^{\prime} \left \langle ~
\hat{n}\left(\mathbf{x}^{\prime}\right)~C(\hat{H})~
\hat{n}\left(\mathbf{x}\right) ~ \right \rangle_{t}
e^{i{\mathbf{Q}} \cdot(\mathbf{x}-\mathbf{x}^{\prime})}.
\end{equation}
Here, $\langle ~\cdots ~\rangle_{t}$ is the expectation value with
respect to the electronic wavepacket at delay-time $t$. The
operator $C(\hat{H}) = \frac{\tau~\Delta E}{\hbar
\sqrt{\pi~8\ln{2}}}\exp(-\frac{\tau^{2}}{8\ln{2}~\hbar^{2}}(\hat{H}-\langle
\hat{H} \rangle_{t})^{2})$ is a function of the electronic
Hamiltonian $\hat{H}$ and of the pulse duration $\tau$. $\Delta E$
refers to the photon energy range accepted by the detector (see
Fig. 2), and $\langle \hat{H} \rangle_{t}$ is the mean energy of
the electronic wavepacket. $\hat{n}(\mathbf{x})$ is the electron
density operator. The expectation value of $\hat{n}(\mathbf{x})$
with respect to the wavepacket at $t$ gives $\rho(\mathbf{x}, t)$
i.e. $\langle \hat{n}(\mathbf{x}) \rangle_{t} = \rho(\mathbf{x},
t)$. As one can see from Eq.~(\ref{eq2}), the quantum description
of light provides a very different expression for the TRI of the
wavepacket, in comparison to that obtained from a classical
description of light, Eq.~(\ref{eq11}). The crucial difference
between both expressions is that Eq.~(\ref{eq11}) depends on
$\langle \hat{n}(\mathbf{x}^{\prime}) \rangle_{t} \langle
\hat{n}(\mathbf{x}) \rangle_{t}$, whereas Eq.~(\ref{eq2}) depends
on $\langle ~
\hat{n}(\mathbf{x}^{\prime})~C(\hat{H})~\hat{n}(\mathbf{x})
~\rangle_{t}$.
The fingerprint of electronic transitions within the finite
spectral bandwidth of the \mbox{X-ray} pulse is encoded in
$C(\hat{H})$, which correlates the scattering of the photon at
different times during the pulse duration. Similar interference
effects caused by interaction of a wavepacket with a photon at
different times during the pulse duration are known, for example,
from fluorescence spectroscopy with optical
lasers~\cite{scherer1991}. Equations (2) and (3) rely on the
assumption that the transverse and longitudinal coherence lengths
of the incident X-ray radiation are larger than the size of the
object.

At this point it is interesting to note that many physical
phenomena involving the interaction of light and matter can be
explained from a classical description of light. Even the
photo-electric effect can be explained by assuming classical
light~\cite{scully1972}, and very sophisticated experimental
scenarios such as in photon antibunching/anticorrelation
experiments~\cite{paul1982, grangier1986} or in Lamb shift
measurements ~\cite{hänsch1972} are necessary to prove the quantum
nature of light~\cite{walls1979}. Surprisingly, a direct
visualization of the quantum nature of light is obtained by
looking at the time-resolved scattering pattern of a
non-stationary quantum system. We illustrate the dramatic
difference between Eqs. (2) and (3) by presenting benchmark
results for TRI of the electronic wavepacket introduced in Fig. 1.
Due to the inherent finite bandwidth of an ultrashort pulse, it is
necessary to compute the transition amplitudes from the electronic
states involved in the wavepacket to all electronic states within
the bandwidth. Depending on the bandwidth, this can cover
transitions to all bound states, including transitions to all
Rydberg states, with any angular momentum. Therefore, the
evaluation of Eq.~(\ref{eq2}) even for hydrogen is numerically
challenging. In contrast, the evaluation of Eq.~(\ref{eq11}) is
easy as it involves only states within the wavepacket.

Figure~3A shows scattering patterns, calculated with
Eq.~(\ref{eq2}), in the $Q_x$ - $Q_z$ plane ($Q_y$ = 0) as a
function of the delay time at times 0, T/4, T/2, 3T/4, and T.
Figure~3B shows the corresponding charge distribution of the
wavepacket, which undergoes spatial oscillations along the
$z$-axis. The probe-pulse duration in Fig.~3A is 1~fs. A photon
energy detection width, $\Delta E$, corresponding to 0.5 eV is
used in the present calculation (see Fig. S1). Therefore, all
transitions induced by the scattering process and resulting in a
scattered photon energy within an energy range of 0.5 eV of the
mean incoming photon energy were considered in the calculation of
the scattering pattern. In order to ensure convergence, we
considered transitions to states up to principal quantum number $n
= 40$. Including all types of multipole transitions allowed by the
conservation of angular momentum, this amounts to about 22000
states.

As the electronic charge  distribution oscillates towards the
negative $z$-axis, the pattern in Fig.~3A oscillates  towards
positive $Q_z$ values. At delay times T/4 and 3T/4, the electronic
charge distributions are identical, whereas the wavepacket carries
a different phase, resulting in different patterns. It is evident
from Fig.~3A that the scattering pattern is not at all times
centrosymmetric i.e. it is not equal for $\mathbf{Q}$ and
$-\mathbf{Q}$. This is in contrast with the centrosymmetric
pattern expected from Eq.~(\ref{eq11}) as a consequence of
Friedel's law~\cite{Nielsen}, and shown in Fig.~3C. The patterns
shown in Fig.~3C were calculated using Eq.~(\ref{eq11}) with a
pulse of duration 1 fs. These patterns are more intense than the
patterns shown in Fig.~3A, in which the intensity depends on the
photon energy detection width.
It is interesting to note that the pixel with maximum intensity
varies as a function of the delay-time in Fig.~3A, whereas in
Fig.~3C it remains fixed at $\mathbf{Q}= 0$.
This counterintuitive nature of the QED scattering pattern arises
from the fact that it is not simply related to the Fourier
transform of the instantaneous electronic density, but it is
related to the electronic wavepacket through Eq.~(\ref{eq2}).
Clearly, the pattern in Fig.~3C is less rich than the correct
pattern in Fig.~3A and it misses important phase information.
Scattering patterns from electronic wavepackets corresponding to
coherent superpositions of different sets of eigenstates (other
than those used in Fig. 1) show similarly dramatic differences
between Eqs.~(\ref{eq11}) and ~(\ref{eq2}). Therefore, Fig.~3A is
an unusually dramatic reflection of the quantum nature of light.

In the illustrative example investigated here, an electronic
wavepacket corresponding to the superposition of two hydrogenic
eigenstates is considered. There have been several studies on the
scattering of quasi-resonant light from strongly driven two-level
systems~\cite{sundaram1990, eberly1992}, particularly under
conditions when the system is prepared initially in a
superposition of states~\cite{gauthey1995}. In those cases the
transitions between electronic states are of dipolar nature. The
TRI regime is, however, in the weak-field, one-photon limit and
the photon energy is highly detuned from the energy differences
between states of the wavepacket. In TRI, scattering events are
Compton-like, i.e., induced by the $\mathbf{\hat{A}}^2$ operator
(see supplementary information), and can couple electronic states
over a wide range of angular momenta. In order to illustrate the
difference between these scenarios, we compute scattering patterns
based on Eq.~(\ref{eq2}) but restrict the calculation to the two
electronic states spanning the prepared wavepacket. The patterns
in Fig. 4 present an intensity similar to that in Fig. 3C because
the same states are involved as in the calculation with
Eq.~(\ref{eq11}). However, a careful look reveals the same
periodicity and non-centrosymmetric features as seen in Fig. 3A.

Coherent diffractive imaging (CDI)~\cite{chapman2010, abbey2011},
which provides access to structural information of
periodic~\cite{zuo2003} and non-periodic~\cite{miao1999} samples,
relies on the Fourier relationship between the scattering pattern
and the electron density. CDI assumes Eq.~(\ref{eq1}) for a
stationary electronic system, implying centrosymmetric patterns as
a consequence of Friedel's law. Consequently, the natural
extension of CDI to the time domain assumes that Eq.~(\ref{eq11})
holds. On the other hand, we see that patterns obtained from a
non-stationary quantum system are not centrosymmetric and cannot
be related via Eq.~(\ref{eq11}) to a real function
$\rho(\mathbf{x},t)$. Therefore, our results demonstrate that the
extension of CDI into the time-domain requires a careful analysis
of the underlying processes.

The scattering patterns calculated from Eqs.~(\ref{eq11}) and
~(\ref{eq2}) are not only unrelated, but the effect of the pulse
duration in the patterns is also totally different.
The respective DSPs for a particular pixel in $Q$-space for
different pulse durations are shown in Fig.~5. The curve
calculated with the correct description, Eq.~(\ref{eq2}), has the
same periodicity as the corresponding wavepacket dynamics
(Fig.~5A). In contrast, the curve calculated with Eq.~(\ref{eq11})
has the wrong periodicity (Fig.~5B). In case of pulses shorter
than the characteristic time scale of the wavepacket, there is no
optimal pulse duration for the signal computed with
Eq.~(\ref{eq11}). The contrast as a function of time increases
monotonically and for short enough pulses becomes almost constant.
Conversely, the pattern computed with Eq.~(\ref{eq2}) has an
optimum contrast as a function of time for a pulse length close to
1~fs. For shorter pulses, the pattern loses contrast again. The
reason behind such an unexpected behavior is that in the case of
hydrogen, the expectation value in Eq.~(\ref{eq2}) becomes
independent of $t$ for $\tau \rightarrow 0$, intuitively because
the electron has no time to change its position during $\tau$.
Consequently, the pattern becomes independent of the electronic
dynamics.

We have shown that TRI from a non-stationary quantum system
provides visual evidence of the quantum nature of light in a very
simple way. Moreover, information on the quantum motion of an
electronic wavepacket beyond the instantaneous electronic density
is accessed by \mbox{X-ray} TRI with atomic-scale spatial and
temporal resolution. Our result is counterintuitive as seen from
the perspective of \mbox{X-ray} scattering from a stationary
electronic system, as one would expect to access the instantaneous
electronic density for sufficiently short pulses. Not only is this
not the case, but the underlying physics fundamentally differs
from such an interpretation. The illustrative example used here as
a proof of principle lies in the time and energy range of interest
corresponding to the dynamics of valence electrons in more complex
molecular and biological systems~\cite{Scholes, Remacle, lenz_jcp,
lenz_jcp1}. We believe that our present findings will help to
solidify the conceptual foundations of the emerging field of TRI,
where understanding the motion of electrons is key to
understanding the functioning of complex molecular and biological
systems. The advent of novel light sources will certainly bring us
closer to this goal.

\begin{acknowledgments}
We thank Sang-Kil Son, Zoltan Jurek and Jyotsana Gupta for useful
discussions.
\end{acknowledgments}

\bibliography{pnas}

\begin{thebibliography}{10}

\bibitem{Ihee}
Ihee H, {et~al.}
\newblock (2005) Ultrafast x-ray diffraction of transient molecular structures
  in solution.
\newblock \emph{Science} 309:1223--1227.

\bibitem{Chapman}
Chapman HN, {et~al.}
\newblock (2011) Femtosecond x-ray protein nanocrystallography.
\newblock \emph{Nature} 470:73--77.

\bibitem{Seibert}
Seibert MM, {et~al.}
\newblock (2011) Single mimivirus particles intercepted and imaged with an
  x-ray laser.
\newblock \emph{Nature} 470:78--81.

\bibitem{ashcroft}
Ashcroft NW, Mermin ND
\newblock (1979) \emph{Solid State Physics}
\newblock (Saunders College, New York).

\bibitem{james1982}
James RW
\newblock (1982) \emph{The Optical Principles of the Diffraction of X-rays}
\newblock (Ox Bow, Woodbridge, Connecticut).

\bibitem{Hauptman}
Hauptman HA
\newblock (1991) The phase problem of x-ray crystallography.
\newblock \emph{Rep. Prog. Phys.} 54:1427--1454.

\bibitem{Miao}
Miao J, Ishikawa T, Shen Q, Earnest T
\newblock (2008) Extending x-ray crystallography to allow the imaging of
  noncrystalline materials, cells, and single protein complexes.
\newblock \emph{Annu. Rev. Phys. Chem.} 59:387--410.

\bibitem{Haessler}
Haessler S, {et~al.}
\newblock (2010) Attosecond imaging of molecular electronic wavepackets.
\newblock \emph{Nature Physics} 6:200--206.

\bibitem{Tzallas}
Tzallas P, Skantzakis E, Nikolopoulos LAA, Tsakiris GD, Charalambidis D
\newblock (2011) Extreme-ultraviolet pump-probe studies of
  one-femtosecond-scale electron dynamics.
\newblock \emph{Nature Physics} 7:781--784.

\bibitem{Hockett}
Hockett P, Bisgaard CZ, Clarkin OJ, Stolow A
\newblock (2011) Time-resolved imaging of purely valence-electron dynamics
  during a chemical reaction.
\newblock \emph{Nature Physics} 7:612--615.

\bibitem{Niikura}
Niikura H, {et~al.}
\newblock (2002) Probing molecular dynamics with attosecond resolution using
  correlated wave packet pairs.
\newblock \emph{Nature} 421:826--829.

\bibitem{Goulielmakis}
Goulielmakis E, {et~al.}
\newblock (2010) Real-time observation of valence electron motion.
\newblock \emph{Nature} 466:739--743.

\bibitem{Sansone}
Sansone G, {et~al.}
\newblock (2010) Electron localization following attosecond molecular
  photoionization.
\newblock \emph{Nature} 465:763--766.

\bibitem{Goulielmakis1}
Goulielmakis E, {et~al.}
\newblock (2008) Single-cycle nonlinear optics.
\newblock \emph{Science} 320:1614--1617.

\bibitem{Hentschel}
Hentschel M, {et~al.}
\newblock (2001) Attosecond metrology.
\newblock \emph{Nature} 414:509--513.

\bibitem{Krausz}
Krausz F, Ivanov M
\newblock (2009) Attosecond physics.
\newblock \emph{Rev. Mod. Phys.} 81:163--234.

\bibitem{Emma}
Emma P, {et~al.}
\newblock (2010) First lasing and operation of an {\aa}ngstrom-wavelength
  free-electron laser.
\newblock \emph{Nature Photonics} 4:641--647.

\bibitem{Pile}
Pile D
\newblock (2011) X-rays: First light from sacla.
\newblock \emph{Nature Photonics} 5:456--457.

\bibitem{Emma1}
Emma P, {et~al.}
\newblock (2004) Femtosecond and subfemtosecond x-ray pulses from a
  self-amplified spontaneous-emission--based free-electron laser.
\newblock \emph{Phys. Rev. Lett.} 92:74801.

\bibitem{Zholents}
Zholents AA, Fawley WM
\newblock (2004) Proposal for intense attosecond radiation from an x-ray
  free-electron laser.
\newblock \emph{Phys. Rev. Lett.} 92:224801.

\bibitem{Starace}
Shao HC, Starace AF
\newblock (2010) Detecting electron motion in atoms and molecules.
\newblock \emph{Phys. Rev. Lett.} 105:263201.

\bibitem{Neutze}
Neutze R, Wouts R, van~der Spoel D, Weckert E, Hajdu J
\newblock (2000) Potential for biomolecular imaging with femtosecond x-ray
  pulses.
\newblock \emph{Nature} 406:752--757.

\bibitem{Quiney}
Quiney HM, Nugent KA
\newblock (2011) Biomolecular imaging and electronic damage using x-ray
  free-electron lasers.
\newblock \emph{Nature Physics} 7:142--146.

\bibitem{Jurek}
Jurek Z, Oszlanyi G, Faigel G
\newblock (2004) Imaging atom clusters by hard x-ray free-electron lasers.
\newblock \emph{Euro. Phys. Lett.} 65:491--497.

\bibitem{Henriksen}
Henriksen NE, Moller KB
\newblock (2008) On the theory of time-resolved x-ray diffraction.
\newblock \emph{J. Phys. Chem. B} 112:558--567.

\bibitem{Tanaka}
Tanaka S, Chernyak V, Mukamel S
\newblock (2001) Time-resolved x-ray spectroscopies: Nonlinear response
  functions and liouville-space pathways.
\newblock \emph{Phys. Rev. A} 63:63405--63419.

\bibitem{scherer1991}
Scherer NF, {et~al.}
\newblock (1991) Fluorescence-detected wave packet interferometry: Time
  resolved molecular spectroscopy with sequences of femtosecond phase-locked
  pulses.
\newblock \emph{J. Chem. Phys.} 95:1487--1511.

\bibitem{scully1972}
Scully MO, Sargent M
\newblock (1972) The concept of the photon.
\newblock \emph{Physics Today} 25:38--47.

\bibitem{paul1982}
Paul H
\newblock (1982) Photon antibunching.
\newblock \emph{Rev. Mod. Phys.} 54:1061--1102.

\bibitem{grangier1986}
Grangier P, Roger G, Aspect A
\newblock (1986) Experimental evidence for a photon anticorrelation effect on a
  beam splitter: a new light on single-photon interferences.
\newblock \emph{Euro. Phys. Lett.} 1:173--179.

\bibitem{hänsch1972}
H{\"a}nsch TW, Shahin IS, Schawlow AL
\newblock (1972) Optical resolution of the lamb shift in atomic hydrogen by
  laser saturation spectroscopy.
\newblock \emph{Nature} 235:63--65.

\bibitem{walls1979}
Walls DF
\newblock (1979) Evidence for the quantum nature of light.
\newblock \emph{Nature} 280:451--454.

\bibitem{Nielsen}
Als-Nielsen J, McMorrow D
\newblock (2011) \emph{Elements of modern X-ray physics}
\newblock (Wiley New York).

\bibitem{sundaram1990}
Sundaram B, Milonni PW
\newblock (1990) High-order harmonic generation: Simplified model and relevance
  of single-atom theories to experiment.
\newblock \emph{Phys. Rev. A} 41:6571--6573.

\bibitem{eberly1992}
Eberly JH, Fedorov MV
\newblock (1992) Spectrum of light scattered coherently or incoherently by a
  collection of atoms.
\newblock \emph{Phys. Rev. A} 45:4706--4712.

\bibitem{gauthey1995}
Gauthey FI, Keitel CH, Knight PL, Maquet A
\newblock (1995) Role of initial coherence in the generation of harmonics and
  sidebands from a strongly driven two-level atom.
\newblock \emph{Phys. Rev. A} 52:525--540.

\bibitem{chapman2010}
Chapman HN, Nugent KA
\newblock (2010) Coherent lensless x-ray imaging.
\newblock \emph{Nature Photonics} 4:833--839.

\bibitem{abbey2011}
Abbey B, {et~al.}
\newblock (2011) Lensless imaging using broadband x-ray sources.
\newblock \emph{Nature Photonics} 5:420--424.

\bibitem{zuo2003}
Zuo JM, Vartanyants I, Gao M, Zhang R, Nagahara LA
\newblock (2003) Atomic resolution imaging of a carbon nanotube from
  diffraction intensities.
\newblock \emph{Science} 300:1419--1421.

\bibitem{miao1999}
Miao J, Charalambous P, Kirz J, Sayre D
\newblock (1999) Extending the methodology of x-ray crystallography to allow
  imaging of micrometre-sized non-crystalline specimens.
\newblock \emph{Nature} 400:342--344.

\bibitem{Scholes}
Scholes GD, Fleming GR, Olaya-Castro A, van Grondelle R
\newblock (2011) Lessons from nature about solar light harvesting.
\newblock \emph{Nature Chemistry} 3:763--774.

\bibitem{Remacle}
Remacle F, Levine RD
\newblock (2006) An electronic time scale in chemistry.
\newblock \emph{Pro. Natl. Acad. Sci. USA} 103:6793--6798.

\bibitem{lenz_jcp}
Dutoi AD, Wormit M, Cederbaum LS
\newblock (2011) Ultrafast charge separation driven by differential particle
  and hole mobilities.
\newblock \emph{J. Chem. Phys.} 134:024303.

\bibitem{lenz_jcp1}
L{\"u}nnemann S, Kuleff AI, Cederbaum LS
\newblock (2008) Charge migration following ionization in systems with
  chromophore-donor and amine-acceptor sites.
\newblock \emph{J. Chem. Phys.} 129:104305.

\end{thebibliography}

\bibliographystyle{pnas2}

\end{article}

\begin{figure*}[ht]
\begin{center}
\centerline{\includegraphics[width=.7\textwidth]{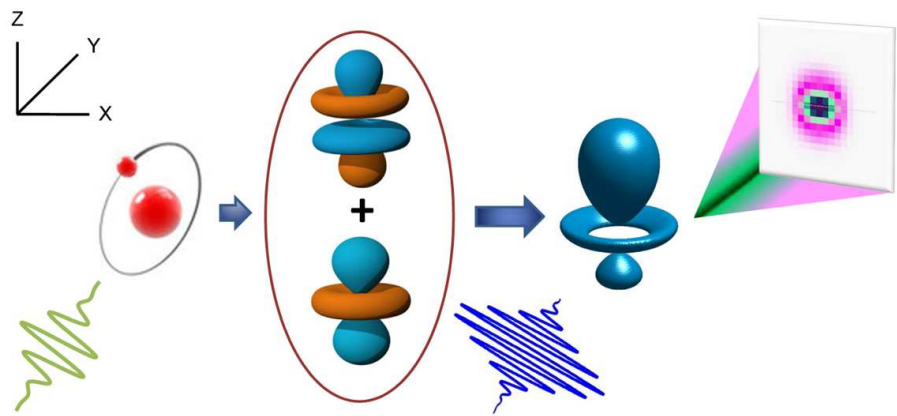}}
\caption{Schematic of the time resolved \mbox{X-ray} imaging
scenario used as an example throughout this work. An electronic
wavepacket is prepared by a laser pump pulse (indicated in green)
as a coherent superposition with equal population of the $3d$ and
$4f$ eigenstates of atomic hydrogen with projection of orbital
angular momentum equal to zero. The polarization direction of the
generated wavepacket is aligned with the laboratory $z$-axis.  The
electronic dynamics of the wavepacket is probed by an ultrashort
\mbox{X-ray} pulse (indicated in blue) propagating along the $y$
direction.  A series of scattering patterns obtained by varying
the pump-probe time-delay serve to image the electronic motion
with high spatial and temporal resolution.}
\end{center}
\end{figure*}

\begin{figure*}[ht]
\begin{center}
\centerline{\includegraphics[width=.4\textwidth]{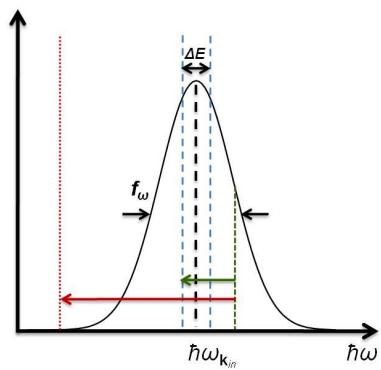}}
\caption{Schematic of the spectral energy distribution of Gaussian
form of an incident ultrashort X-ray pulse with spectral bandwidth
$f_{\omega}$ centered at the incident photon energy $\hbar
\omega_{\mathbf{k}_{in}}$ (see supplementary information). The
window function $W_{\Delta E}({\hbar \omega_{\mathbf{k}_{s}}})$ is
a function of the scattered photon energy $\hbar
\omega_{\mathbf{k}_{s}}$ with a width $\Delta E$ (indicated by
dashed blue lines). $\Delta E$ models the range of photon energies
accepted by the detector. Therefore, elastic or inelastic
transitions taking place during the scattering process that fall
inside $\Delta E$ are considered in the evaluation of the
scattering pattern (indicated by a green arrow). Transitions
falling outside the range of $\Delta E$ are not considered in the
evaluation of the scattering pattern (indicated by a red arrow).}
\end{center}
\end{figure*}

\begin{figure*}[ht]
\begin{center}
\centerline{\includegraphics[width=.6\textwidth]{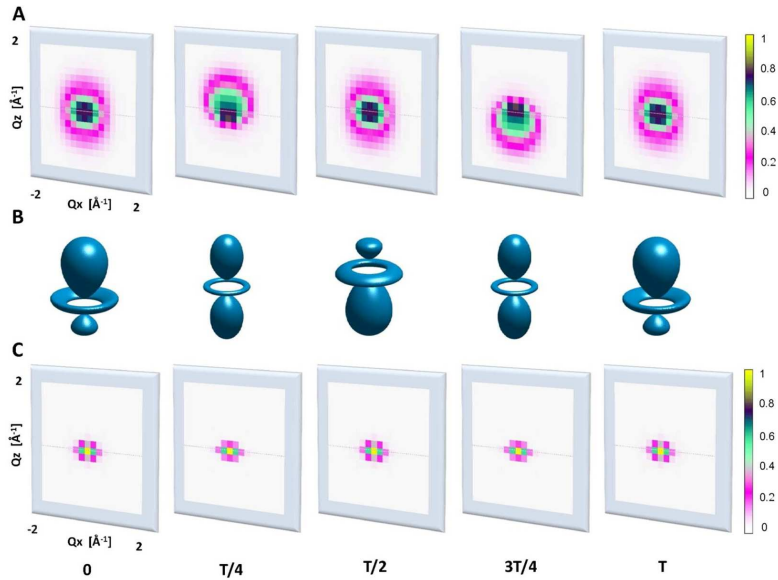}}
\caption{Scattering patterns in the $Q{_x}$ - $Q{_z}$ plane
($Q{_y}$ = 0) and electronic charge distributions of the
wavepacket corresponding to the coherent superposition of $3d$ and
$4f$ eigenstates of atomic hydrogen (cf. Fig. 1). (A) Scattering
patterns obtained with Eq.~(\ref{eq2}), (B) electronic charge
distributions and (C) scattering patterns obtained with
Eq.~(\ref{eq11}), at pump-probe delay times 0, T/4, T/2, 3T/4, and
T, where the oscillation period of the electronic wavepacket is T
= 6.25 fs. The isosurface in (B) encloses $\sim 26 \%$ of the
total probability and has length 14--17 \AA~ along the $z$-axis
and 7.5--9 \AA~ along the $x$ and $y$-axes. The wavepacket is
exposed to a 1 fs \mbox{X-ray} pulse with 4~keV photons. The
patterns are calculated for ${Q}_{\mathrm{max}} = 2$ \AA$^{-1}$
corresponding to 3.14 \AA ~spatial resolution. For the pulse used,
a real-space pixel size of 3.14 \AA ~requires the detection of
photons scattered up to 60$^{\circ}$ with respect to the
\mbox{X-ray} propagation direction, which is the $y$-axis in the
present case. The grid spacing in $Q$-space is 0.19 \AA$^{-1}$.
The intensities of the patterns are shown in units of
$\frac{dP_{e}}{d\Omega}$ in both cases.}
\end{center}
\end{figure*}

\begin{figure*}[ht]
\begin{center}
\centerline{\includegraphics[width=.7\textwidth]{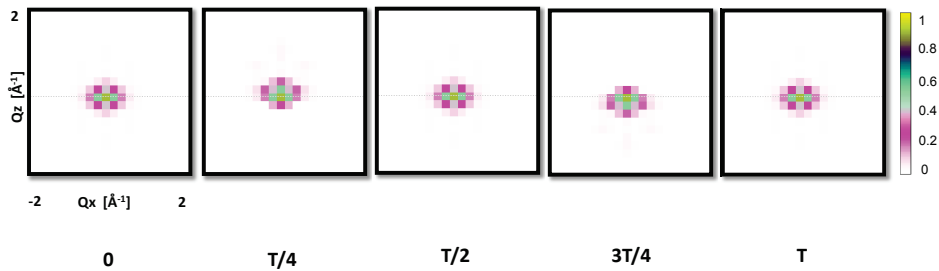}}
\caption{Scattering patterns in the $Q{_x}$ - $Q{_z}$ plane
($Q{_y}$ = 0) obtained with Eq.~(\ref{eq2}) at pump-probe delay
times 0, T/4, T/2, 3T/4, and T, where the oscillation period of
the electronic wavepacket is T = 6.25 fs. All the parameters used
to obtain the patterns are the same as used in Fig. 3A. However,
only the eigenstates belonging to the electronic wavepacket ($3d$
and $4f$ eigenstates of atomic hydrogen with angular momentum
projection quantum number equal to zero) were considered. The
intensities of the patterns are shown in units of
$\frac{dP_{e}}{d\Omega}$.}
\end{center}
\end{figure*}

\begin{figure*}[ht]
\begin{center}
\centerline{\includegraphics[width=.4\textwidth]{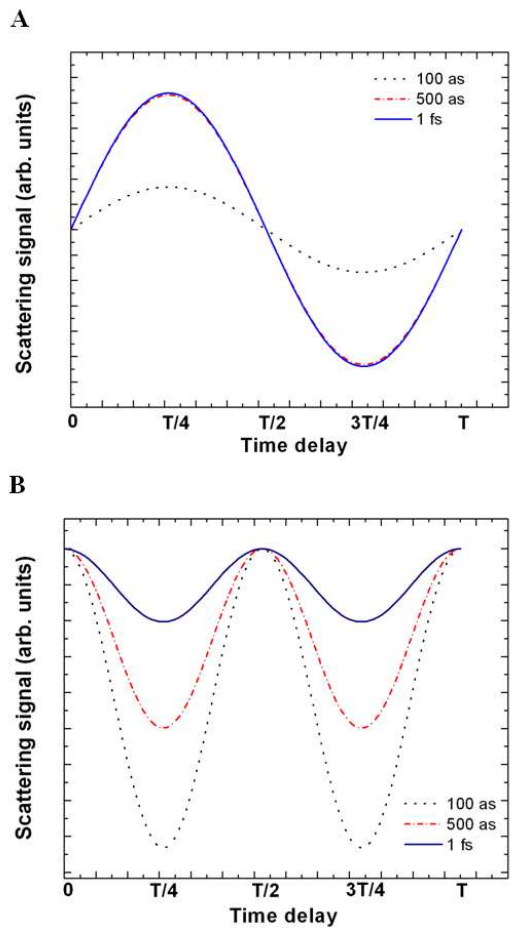}}
\caption{Effect of duration (full width at half maximum) of the
\mbox{X-ray} pulse on the scattering patterns as a function of the
pump-probe delay time. The electronic wavepacket oscillation
period is T = 6.25 fs. The time evolution of the scattering
pattern at an individual pixel as a function of the delay time is
calculated using (A) Eq.~(\ref{eq2}), and (B) Eq.~(\ref{eq11}),
for different pulse durations, 100 as (black dots), 500 as (red
dash-dots), and 1 fs (blue line). In the case of the pattern
calculated using Eq.~(\ref{eq2}), there is an optimal pulse
duration of about 1~fs, which provides a maximum contrast as a
function of time.  There is no such optimum for the pattern
computed with the simple expression in Eq.~(\ref{eq11}).}
\end{center}
\end{figure*}

\end{document}